\documentclass[aps,prl,twocolumn,superscriptaddress]{revtex4-2}

\usepackage{amsmath}
\usepackage{amssymb}
\usepackage{amsfonts}
\usepackage{graphicx}
\usepackage{hyperref}

\begin{document}


\title{Weak Chaos, Anomalous Diffusion,\\and Weak Ergodicity Breaking in Systems with Delay}


\author{Tony Albers}
\email[]{tony.albers@physik.tu-chemnitz.de}
\author{Lukas Hille}
\email[]{hille.lukas@proton.me}
\author{David M\"uller-Bender}
\email[]{david.mueller-bender@mailbox.org}
\affiliation{Institute of Physics, Chemnitz University of Technology, 09107 Chemnitz, Germany}
\author{G\"unter Radons}
\email[]{radons@physik.tu-chemnitz.de}
\affiliation{Institute of Physics, Chemnitz University of Technology, 09107 Chemnitz, Germany}
\affiliation{ICM - Institute for Mechanical and Industrial Engineering, 09117 Chemnitz, Germany}


\date{\today}

\begin{abstract}
We demonstrate that standard delay systems with a linear instantaneous and a delayed nonlinear term show weak chaos, asymptotically subdiffusive behavior, and weak ergodicity breaking
if the nonlinearity is chosen from a specific class of functions.
In the limit of large constant delay times, anomalous behavior may not be observable due to exponentially large crossover times.
A periodic modulation of the delay causes a strong reduction of the effective dimension of the chaotic phases, leads to hitherto unknown types of solutions,
and the occurrence of anomalous diffusion already at short times.
The observed anomalous behavior is caused by non-hyperbolic fixed points in function space.
\end{abstract}


\maketitle

Anomalous diffusion is a frequently observed transport mechanism, where the mean-squared displacement increases nonlinearly in time \cite{klages2008}.
It can be found in physical, biological, and chemical systems such as charge-carrier transport in amorphous semiconductors \cite{scher1975},
the motion of macromolecules in the cytoplasm of living cells \cite{weiss2004}, and the dynamics of monomers belonging to long polymers \cite{shusterman2004}.
While anomalous diffusion is often described by stochastic models such as fractional Brownian motion \cite{mandelbrot1968} or continuous-time random walks \cite{metzler2000},
it was also found in purely deterministic systems due to the presence of chaos.
This so-called anomalous chaotic diffusion has been investigated intensively, but typically only for low-dimensional systems such as iterated maps \cite{geisel1984,geisel1985,zumofen1993,bel2006}
or Hamiltonian systems \cite{chirikov1979,lichtenberg1992,zacherl1986,geisel1987,zumofen1994}.
There are only few works dealing with chaotic diffusion in higher-dimensional or even infinite-dimensional systems
whose dynamics is ruled by delay differential equations \cite{wischert1994,schanz2003,sprott2007,lei2011,dao2013_1,dao2013_2,mackey2021,albers2022_1,albers2022_2}
or partial differential equations \cite{cisternas2016,cisternas2018,albers2019_1,albers2019_2}.
Except for two cases in the latter class \cite{cisternas2016,cisternas2018}, the diffusion there was always normal.
In this Letter, we show that anomalous diffusion can also be found in a standard family of systems with delay.
Furthermore, the observed dynamics shows weak chaos \cite{korabel2009,korabel2010}, i.e., the maximal Lyapunov exponent vanishes.
\footnote{Note that the term weak chaos is also used with a different meaning in the delay literature \cite{heiligenthal2011}}.
We also find weak ergodicity breaking \cite{bouchaud1992,lubelski2008,he2008} implying an inequivalence of ensemble and time averages.
To the best of our knowledge, such phenomena have never been observed before in formally infinite-dimensional systems described by delay differential equations,
which are important because they have many applications in physics, biology, and engineering \cite{stepan1989,michiels2007,erneux2009,lakshmanan2011}.
Concrete examples are chaos control \cite{schoell2007}, population dynamics \cite{kuang1993}, turning and milling processes \cite{otto2015}, and nonlinear optics \cite{ikeda1980,mueller2020}.

We consider a typical class of time-delayed feedback systems with linear instantaneous and nonlinear delayed term described by the delay differential equation (DDE)
\begin{equation}
\label{eq:DDE}
\frac{1}{\Theta}\,\dot{x}(t)=-x(t)+f\boldsymbol{(}x[R(t)]\boldsymbol{)}\,.
\end{equation}
The parameter $\Theta$ sets the overall timescale, where the case $\Theta\gg1$ corresponds to the large delay limit and Eq.~(\ref{eq:DDE}) then belongs to the class of singularly perturbed DDEs
\cite{ikeda1982,chow1983,mallet-paret1986,ikeda1987,mensour1998,wolfrum2006,adhikari2008,wolfrum2010,lichtner2011,giacomelli2012,marino2014,amil2015,faggian2018,marino2019}.
The retarded argument $R(t)=t-\tau(t)$ is determined by a time-varying delay $\tau(t)$, where we choose a periodic delay modulation $\tau(t)=\tau_0-A/(2\pi)\sin(2\pi t)$ with mean delay $\tau_0$ and amplitude $A$.
For $A=0$, one recovers the well-studied case of constant delay $\tau=\tau_0$ \cite{farmer1982,ikeda1982,chow1983,mallet-paret1986,ikeda1987,mensour1998,adhikari2008,amil2015}.
Eq.~(\ref{eq:DDE}) includes famous examples for specific choices of the nonlinearity $f(x)$ such as the Mackey-Glass equation \cite{mackey1977} and the Ikeda equation \cite{ikeda1980,ikeda1982}.
In the following, we consider a certain class of nonlinearities such that the resulting DDE has unstable hyperbolic fixed point solutions (HFPS),
but also unstable nonhyperbolic fixed point solutions (NHFPS) in function space.
For the numerical investigations of Eq.~(\ref{eq:DDE}), we use as a specific example the so-called double-sine nonlinearity
\begin{equation}
\label{eq:double_sine}
f_{\mu}(x)=x-\mu\left(\sin(2\pi x)+\frac{1}{2}\sin(4\pi x)\right)\,,
\end{equation}
which is a natural extension of the previously studied climbing-sine nonlinearity with only unstable HFPS in the corresponding DDE which results in normal chaotic diffusion \cite{albers2022_1}.
It has reflection symmetry $f_{\mu}(-x)=-f_{\mu}(x)$ and discrete translational symmetry $f_{\mu}(x+1)=f_{\mu}(x)+1$, where the latter allows to divide the coordinate axis into unit cells.
Eq.~(\ref{eq:DDE}) with Eq.~(\ref{eq:double_sine}) possesses a lattice of NHFPS $x_{\text{nh}}^*(t)=1/2+n\in\mathbb{Z}$ at the half integers and HFPS $x_{\text{h}}^*(t)=n\in\mathbb{Z}$ at the integers.
A corresponding linear stability analysis of the DDE shows that for the NHFPS one stability exponent is equal to zero while all the others have negative real part \cite{gushchin1999}.
For the HFPS, all stability exponents are complex with a finite number having positive real part while all others have negative real part.
For constant delay, one knows in addition that these eigenvalues form a pseudo-continuous spectrum \cite{lichtner2011,yanchuk2015,yanchuk2017}.
We will demonstrate that the resulting time-delayed system, Eq.~(\ref{eq:DDE}) with Eq.~(\ref{eq:double_sine}), can show weak chaos, chaotic subdiffusion, and weak ergodicity breaking,
where the variable $x(t)$ can be interpreted as an unbounded phase variable.
We point out, however, that the problem is formally infinite-dimensional because Eq.~(\ref{eq:DDE}) can be solved by the method of steps \cite{bellman1965}
leading to an iteration of functions, the so-called solution segments $x_n(t)$ defined on state intervals $(t_{n-1},t_n]$ with $t_{n-1}=R(t_n)$.
These solution segments can show diffusion but because they are of finite variation, it is sufficient for diffusion studies to consider only one value of each segment.

\begin{figure}
\includegraphics[width=\linewidth]{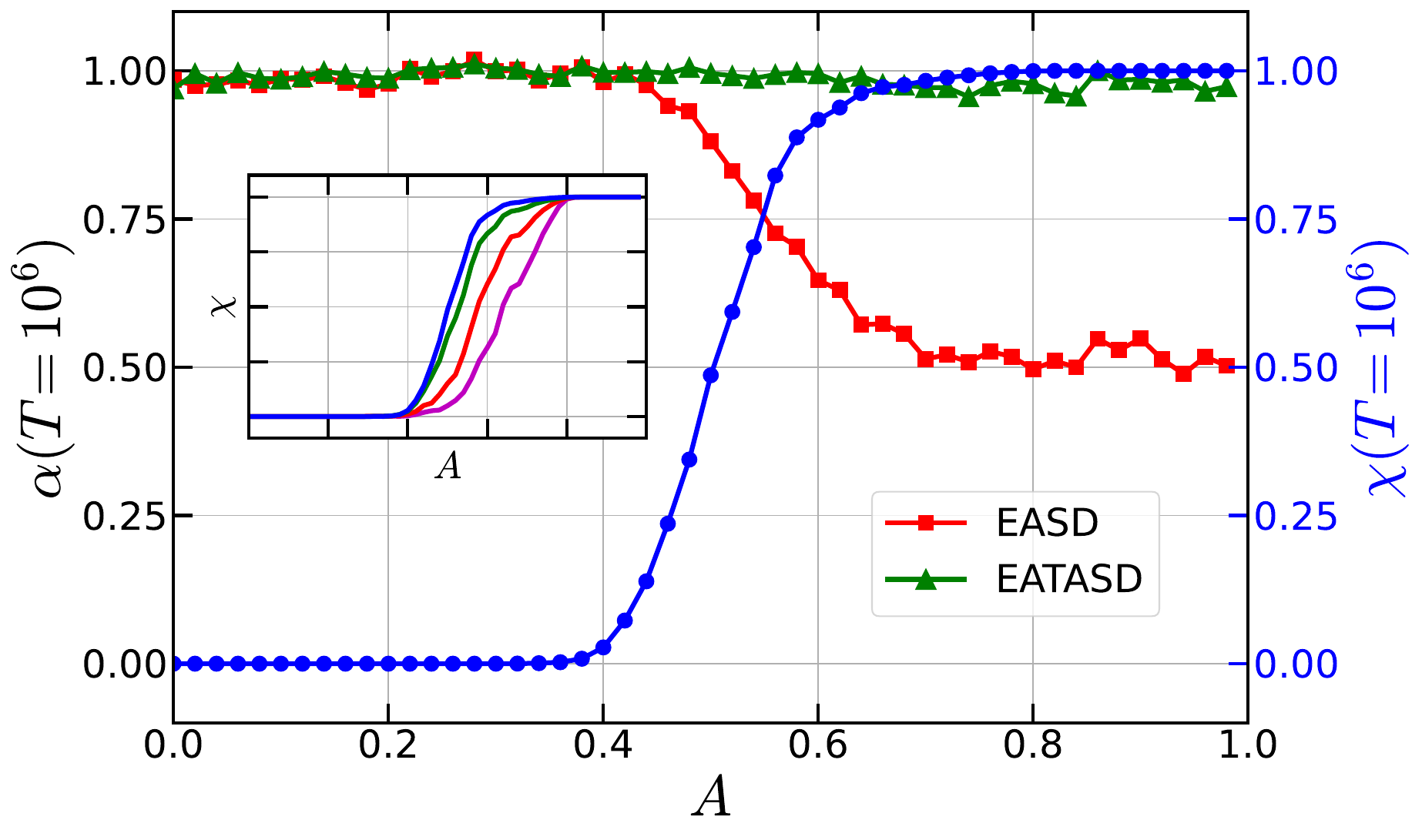}
\caption{\label{fig:1}
The diffusion exponent $\alpha$ from the EASD (red squares) shows an apparent transition from normal diffusion ($\alpha=1$) to subdiffusion ($\alpha=1/2$) when the modulation amplitude $A$ increases from zero to one.
The diffusion exponent from the EATASD (green triangles) is always equal to one indicating weak ergodicity breaking.
This is related to the mean fraction $\chi$ of plateau phases (blue circles), which increases from zero (only turbulent phases, see Fig.~\ref{fig:2} left or right) to one (only plateau phases).
Because of the nonstationarity of the system, all these results depend on the observation time $T$ as shown in the inset for the fraction $\chi$ ($T=10^3,10^4,10^5,10^6$ from bottom to top, ensemble size $N=1024$).}
\end{figure}

We characterize the observed diffusion on the one hand by the ensemble-averaged squared displacement (EASD) $\langle\Delta x^2(t)\rangle_{\text{E}}:=\langle[x(t)-x(0)]^2\rangle_{\text{E}}\sim t^{\alpha}$,
which typically increases according to a power law with the diffusion exponent $\alpha$,
and on the other hand by the time-averaged squared displacement (TASD) $\langle\Delta x^2(t)\rangle_{\text{T}}:=1/(T-t)\int_0^{T-t}[x(t'+t)-x(t')]^2\,\text{d}t'$.
Because the latter can be a random variable even for $T\to\infty$, we consider its ensemble average (EATASD).
To study the influence of the time-varying delay, we fix the mean delay, $\tau_0=1$,
and we vary the amplitude of the delay modulation in the range $0\leq A\leq1$ such that the retarded argument $R(t)$ remains invertible.
Furthermore, we set $\Theta=50$ for all figures because we are interested in the large delay limit.
While in Fig.~\ref{fig:1} the numerically observed diffusion exponent from the EASD varies between 1 and 1/2,
the exponent from the EATASD is found to be one for every $A$ indicating a nonequivalence of ensemble and time averages.
This is the first time that anomalous diffusion and weak ergodicity breaking is found in a delay system.

\begin{figure}
\includegraphics[width=\linewidth]{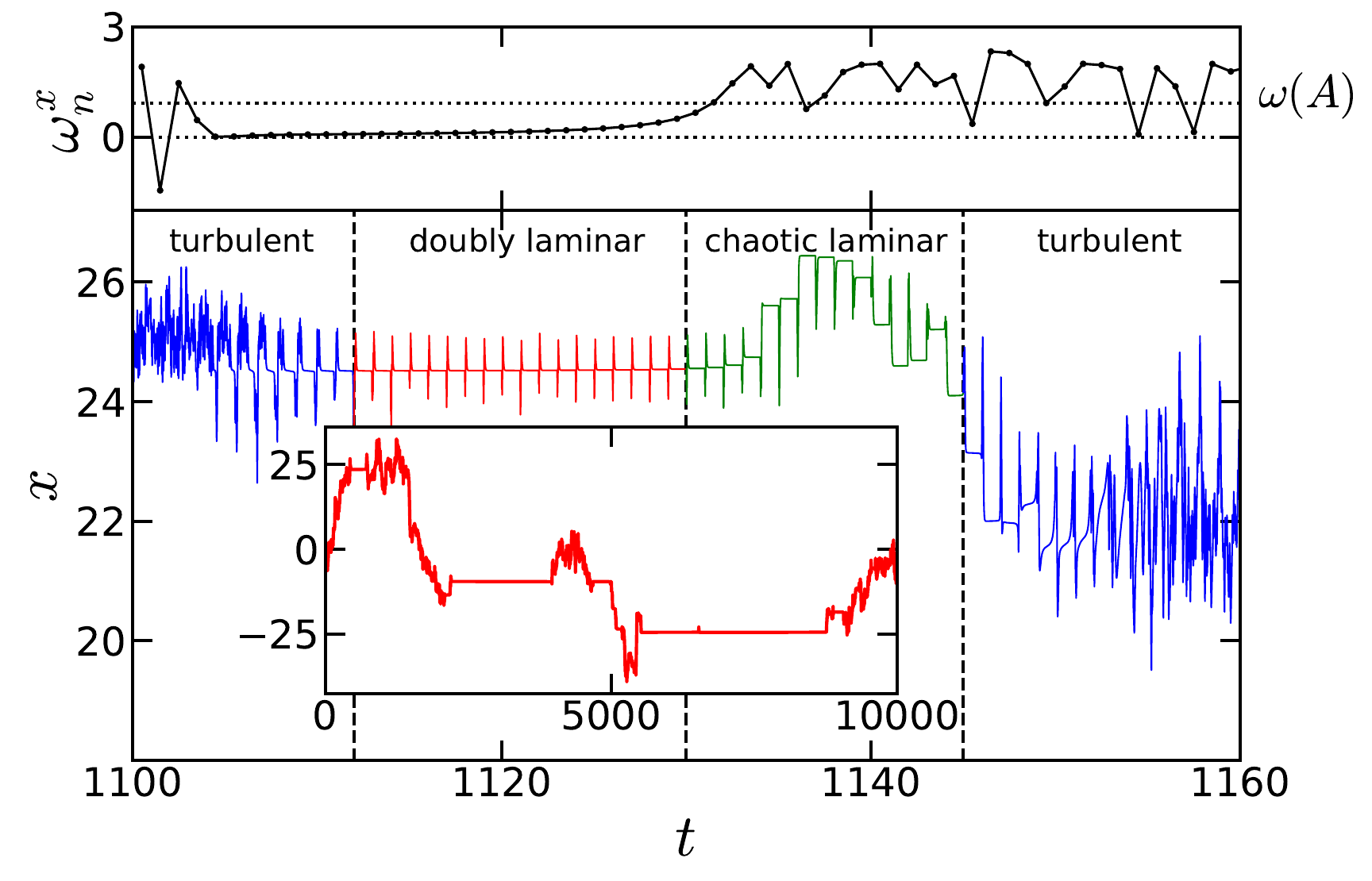}
\caption{\label{fig:2}
A single solution $x(t)$ of the DDE on a short timescale for $A=0.6$ shows a coexistence of doubly laminar, chaotic laminar, and turbulent phases ($\Theta=50$, $\tau_0=1$, and $\mu=0.9$).
Plateau phases begin (end) when the instantaneous expansion rate $\omega_n^x$ in coordinate direction is smaller (larger) than the expansion rate $\omega(A)$ on the time axis for a longer phase.
The inset shows the same solution on a larger timescale, which due to its long waiting times (doubly laminar phases) is reminiscent of a subdiffusive continuous-time random walk.}
\end{figure}

\begin{figure}
\includegraphics[width=\linewidth]{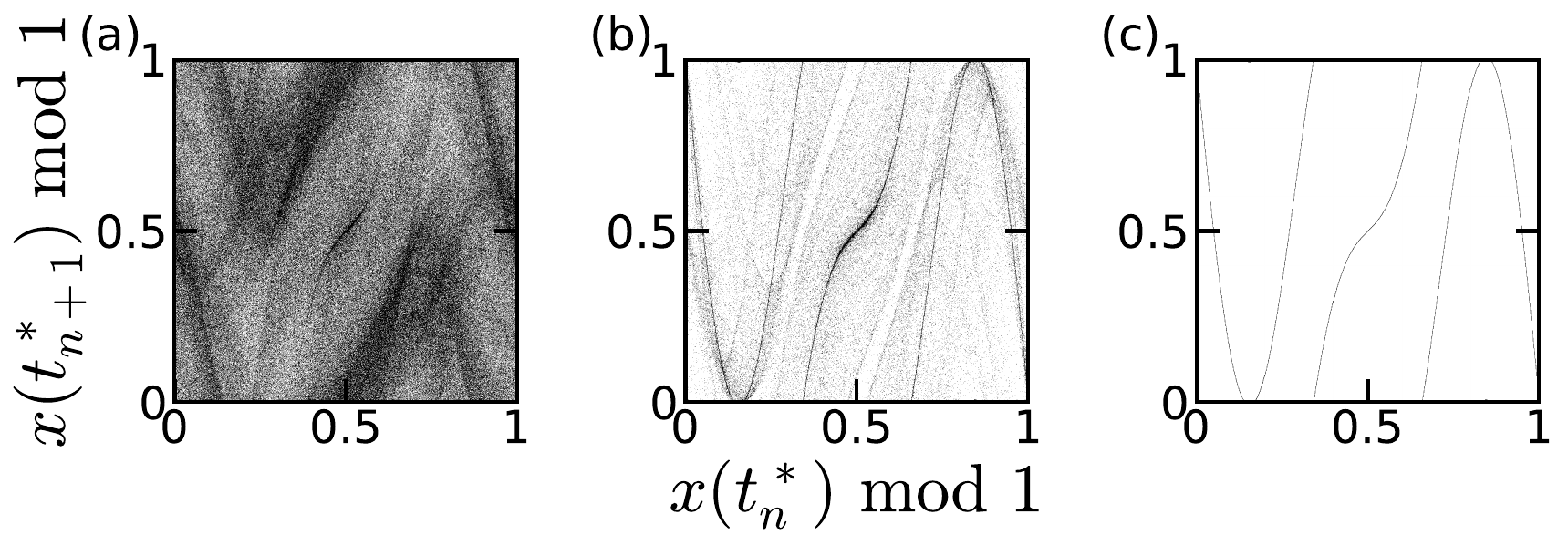}
\caption{\label{fig:4}
Return maps for different values of the modulation amplitude $A$ ($A=0.5,0.7,0.9$ from left to right).
For larger values of $A$, one recognizes the structure of the double-sine map $x_{n+1}=f_{\mu}(x_n)$, but for decreasing values of $A$, the picture becomes more and more blurred
due to high-dimensional chaotic bursts (see Fig.~\ref{fig:5}).}
\end{figure}

To understand the transition from normal to anomalous diffusion, we first consider a single solution of the DDE, which was obtained numerically
\footnote{The DDE in Eq.~(\ref{eq:DDE}) was solved numerically using the two-stage Lobatto IIIC method \cite{bellen2003} with linear interpolation and a step width $\Delta t=0.0001$.}
for an intermediate value of the modulation amplitude $A$ (see Fig.~\ref{fig:2}).
On a short time scale (main figure), this time series shows a switching between two types of laminar and turbulent phases
\footnote{
A solution $x_n(t)$ in one state interval $(t_{n-1},t_n]$ is identified as a plateau if the half of all derivatives $\dot{x}_n(t)$ of the solution are smaller than a certain threshold value $d_c=0.001$,
i.e., if the median $q_{1/2}$ of all derivatives in one state interval is smaller than $d_c$.
A plateau phase is identified as a doubly-laminar phase if the value of the plateau $x_n=x_n(t^*)$ fulfills the condition $|x_n-x^*|<0.05$.}.
Turbulent phases are characterized by chaotic high frequency fluctuations similar to turbulent chaos \cite{ikeda1980} as known from systems with constant delay ($A=0$).
Both forms of the laminar phases are characterized by a sequence of nearly constant plateaus (interrupted by irregular bursts).
Therefore, we refer to them collectively as plateau phases.
The special case of chaotic laminar phases has recently been discovered \cite{mueller2018,mueller2019}
and investigated experimentally \cite{hart2019,mueller2020,juengling2020,kulminskii2020} for delay systems as in Eq.~(\ref{eq:DDE}).
They are called laminar due to the laminar motion within the plateaus, and they are chaotic because the heights of the plateaus are mapped from one plateau to the next by the 1-d map $x'=f(x)$
with $f(x)$ from Eq.~(\ref{eq:double_sine}), which is chaotic if its Lyapunov exponent is positive $\lambda_f>0$.
For unbounded motion of the plateaus this can lead to (normal) laminar chaotic diffusion \cite{albers2022_1}.
The condition for the development of plateaus is $\lambda_f+\lambda_R<0$, where $\lambda_R$ is the Lyapunov exponent of the 1-d map $t'=R(t)$.
This inequality just expresses that the stretching of the signal $x_n(t)$ in the $t$-direction (due to the action of the Koopman operator associated with $R(t)$) in each step is on average stronger
than the enlargement of fluctuations of $x_n(t)$ (due to the action of the nonlinear operator associated with $f(x)$).
For details see \cite{mueller2018,mueller2019}.
The second kind of our plateau phases is characterized by plateaus that hardly change their heights when mapped from one state interval to the next.
They are laminar in a double sense, within each plateau and from one plateau to the next, and therefore called doubly laminar.
The latter motion is very regular and non-chaotic, and plays a dominant role for the long-term dynamics of the delay system.
Their appearance can be understood by noting that the map $x'=f_{\mu}(x)$ (due to Eq.~(\ref{eq:double_sine}) called double-sine map), belongs to
the well-studied class of Pomeau-Manneville maps \cite{manneville1979,pomeau1980,manneville1980} because it has marginally unstable fixed-points at $x^*=1/2+n\in\mathbb{Z}$
where the map for $x\rightarrow x^*$ behaves as $f_{\mu}(x)\simeq x+4\pi^3\mu(x-x^*)^3$.
According to the theory in \cite{geisel1984}, this leads to long residence times $\Delta$ close to $x^*$ with probability distribution $\psi(\delta)\sim\delta^{-3/2}$.
Hence, the mean residence time close to $x^*$, where $f_{\mu}'(x^*)=1$, diverges.
As a consequence, the standard Lyapunov exponent $\lambda_f$ of the map in the long time limit is equal to zero, i.e., the map shows weak chaos \cite{korabel2009,korabel2010}.
Furthermore, for our parameter ranges, $\lambda_R=\ln(1-A)$ is always negative except for constant delay ($A=0$), where $\lambda_R=0$,
such that the above mentioned condition, extended to weak laminar chaos, is fulfilled for every $0<A<1$.
However, such Pomeau-Manneville like maps are known to show weak ergodicity breaking typically implying strong sample to sample fluctuations of time averaged quantities
such as the Lyapunov exponent \cite{korabel2009,korabel2010} even in the long-time limit and especially on a shorter time scale resulting in the switching behavior in Fig.~\ref{fig:2}.
In more detail, we assign to each solution segment $x_n(t)$ taken at the attractive fixed points $t_n^*$ of $R(t)+1$ an instantaneous expansion rate $\omega_n^x=\ln|f_{\mu}'(x_n(t_n^*))|$
in coordinate direction and an instantaneous expansion rate on the time axis $\omega_n^t=\ln|(R^{-1})'(t_n^*)|=-\ln|R'(t_n^*)|=-\ln(1-A)=\omega(A)$, which is independent of the number $n$ of the segment $x_n(t)$.
Plateaus can develop if $\omega_n^x<\omega(A)$ for a longer period.
This is the case if $x_n(t_n^*)$ comes close to the marginally unstable fixed points of the double-sine map (see Fig.~\ref{fig:2}).
For laminar chaos, the dynamics of the plateau heights is governed by the iterated map $x'=f_{\mu}(x)$ (see Fig.~\ref{fig:4}(c)),
which in our case shows intermittency with long residence times close to the marginally unstable fixed points (resulting in the delay system in the doubly laminar phases)
interrupted by chaotic bursts (resulting in chaotic laminar phases).
During the latter, $\omega_n^x$ can be larger than $\omega(A)$ for a longer period leading to a destruction of the chaotic laminar phase and to the beginning of a new turbulent phase.
This switching between doubly laminar, chaotic laminar, and turbulent phases within one single time series is a new type of dynamical behavior. 

\begin{figure}
\includegraphics[width=\linewidth]{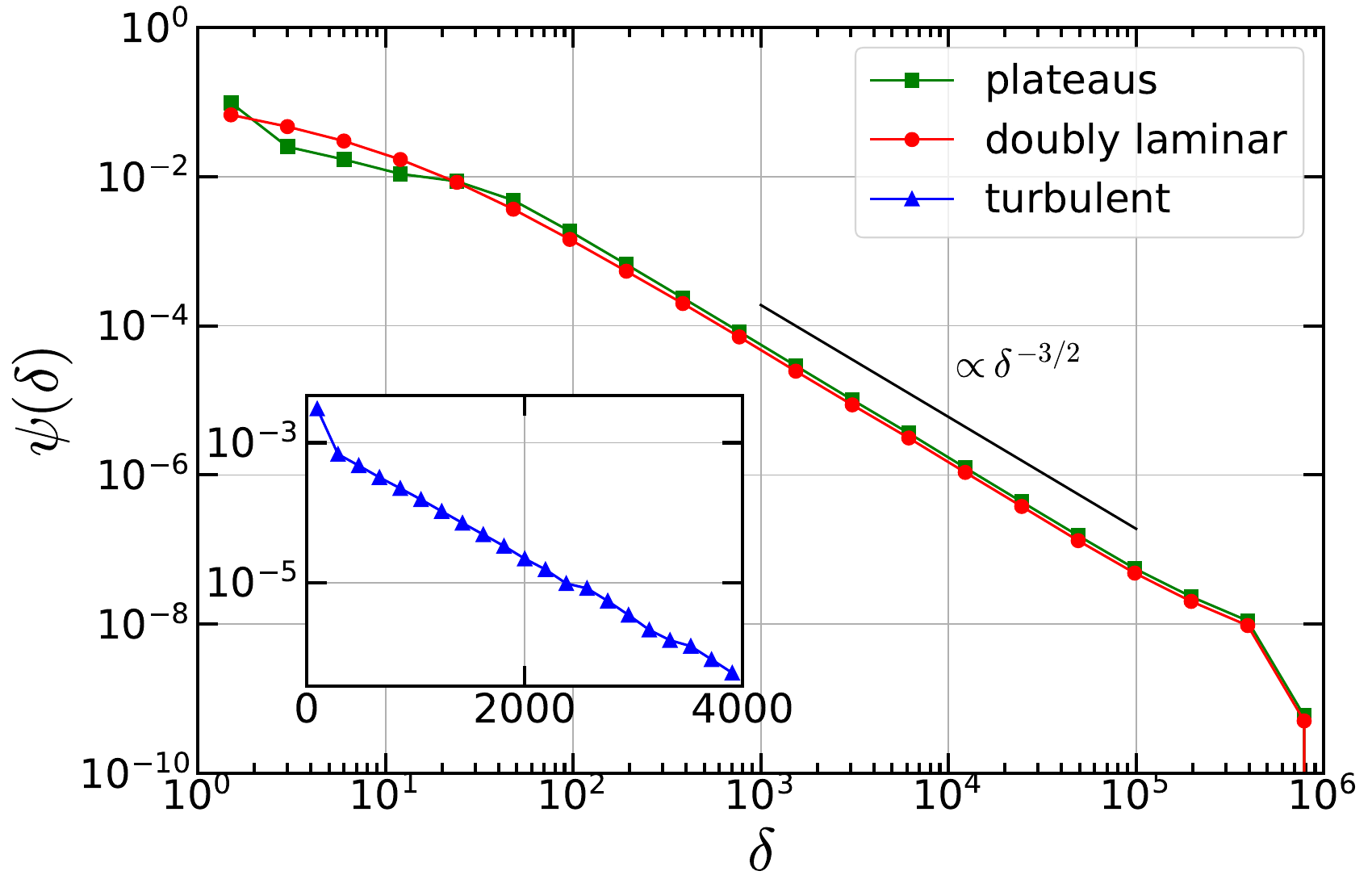}
\caption{\label{fig:3}
The distributions $\psi_{\text{l}}(\delta)$ of durations of plateau phases and $\psi_{\text{dl}}(\delta)$ of doubly laminar phases are both heavy tailed decaying asymptotically as $\delta^{-3/2}$.
In contrast, the inset shows that the distribution $\psi_{\text{t}}(\delta)$ of the durations of turbulent phases is exponential (same parameters as in Fig.~\ref{fig:2}).}
\end{figure}

Next, we characterize this behavior statistically.
First, we consider the distributions of durations of turbulent and plateau phases for one value of the modulation amplitude $A$ (see Fig.~\ref{fig:3}).
The durations of the doubly laminar phases are found to be asymptotically equal in distribution to the residence times close to the marginally unstable fixed points of the double-sine map.
Therefore, the distribution $\psi_{\text{dl}}(\delta)$ of doubly laminar phases is heavy tailed and decays asymptotically as $\delta^{-3/2}$.
Because plateau phases consist of doubly laminar phases and chaotic laminar phases, where the latter are exponentially distributed, doubly laminar and plateau phases show the same heavy-tails in their distributions.
The distribution of the durations of turbulent phases is also exponential.
These basic properties appear to hold for all values of $A$, even for $A=0$.

\begin{figure}
\includegraphics[width=\linewidth]{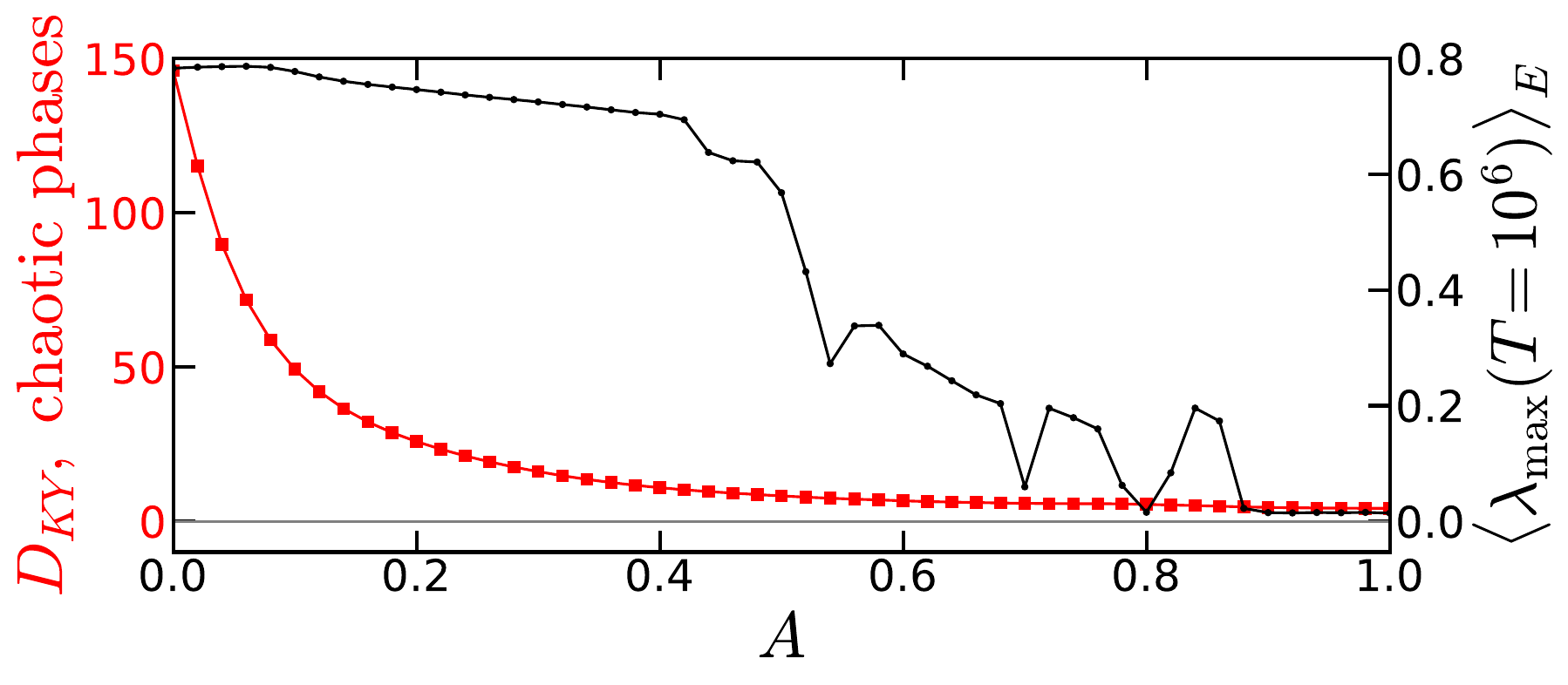}
\caption{\label{fig:5}
The Kaplan-Yorke dimension $D_{\text{KY}}$ calculated only from the chaotic (turbulent and chaotic laminar) phases shows a strong reduction of their effective dimension as the modulation amplitude $A$ is increased.
According to the fraction $\chi$ of plateau phases shown in Fig.~\ref{fig:1}, the maximal finite-time Lyapunov exponent $\langle\lambda_{\text{max}}(T)\rangle_{\text{E}}$ is nearly zero for ranges of $A$,
where only plateau phases are present.}
\end{figure}

A variation of the modulation amplitude $A$, however, drastically changes the parameters of the distributions, which has a strong impact on the diffusive behavior of the system.
To understand this in more detail, we first consider the mean fraction $\chi$ of plateau phases, i.e.,
the sum of all durations of chaotic laminar and doubly laminar phases in one single time series divided by the length of the time series averaged over an ensemble of trajectories (see Fig.~\ref{fig:1}).
This fraction strongly depends on the modulation amplitude $A$ and increases from almost zero (effectively only turbulent phases) to one (only plateau phases) as $A$ is varied from zero to one (main figure).
Because of the diverging mean of the distribution of plateau phases and the resulting nonstationarity, the fraction $\chi=\chi(T)$ depends on the observation time $T$ (see inset).
This fraction is also reflected in the return maps in Fig.~\ref{fig:4}, which show how the solution taken in the middle of a state interval is correlated with the corresponding value from the next state interval.
For larger values of $A$, where (weak) laminar chaos prevails, one can see the structure of the double-sine map which becomes successively blurred for decreasing values of $A$.
The blurring is caused by high-dimensional excursions of the state function (see Fig.~\ref{fig:5}).
The fraction of plateau phases is related to the diffusive properties of the system as follows.
We numerically found that turbulent phases always show normal chaotic diffusion, whereas the plateaus follow the dynamics of the double-sine map.
This map shows chaotic diffusion due to a ``transfer range'' in each unit cell to one of the neighboring cells \cite{geisel1982},
which becomes subdiffusive due to the long residence times close to the marginally unstable fixed points \cite{geisel1984}.
It is known that such kind of dynamics can be described statistically by a subdiffusive continuous-time random walk with instantaneous jumps and heavy-tailed waiting time distributions \cite{zumofen1993}.
The EASD asymptotically increases according to a power law, i.e., $\langle\Delta x^2(n)\rangle_{\text{E}}\sim n^{1/2}$,
whereas the EATASD grows linearly in time which is a manifestation of weak ergodicity breaking \cite{lubelski2008,he2008}.
This explains why normal diffusion is observed in Fig.~\ref{fig:1} if $\chi\approx0$, whereas subdiffusion is observed in the EASD if $\chi\approx1$ while the EATASD always shows normal diffusion.
Furthermore, our time-delayed system shows weak chaos in the sense that the maximal Lyapunov exponent $\lambda_{\text{max}}(T=\infty)$ vanishes
\footnote{The Lyapunov exponents of DDE~(\ref{eq:DDE}) were computed using the method in \cite{farmer1982},
where an integration step width $\Delta t=0.001$ was used and the separation functions governed by the linearized system were reorthonormalized after each delay period.
For the computation of $D_{KY}$ of the chaotic phases, the Lyapunov exponents were defined by the time average of the expansion rates inside the chaotic laminar and turbulent phases,
excluding the doubly laminar phases.}
\footnote{The irregular behavior of $\lambda_{\text{max}}(T)$ for intermediate values of A can be explained by the dynamics of the bursts in the doubly laminar phases (see Fig.~\ref{fig:2}),
where stable (nearly) periodic and chaotic burst dynamics is observed.
Chaotic burst dynamics is associated with positive expansion rates and thus increases the overall Lyapunov exponent in contrast to stable periodic burst dynamics.}.
For finite observation times $T$, $\lambda_{\text{max}}(T)$ also exhibits the apparent transition between normal (positive) and anomalous behavior (zero) as shown in Fig.~\ref{fig:5}.

Also for the extreme case of constant delay, $A=0$, the occurrence of anomalous diffusion and weak chaos can be explained by an analytical approach
using the existence of NHFPS in function space whose largest stability exponent is equal to zero.
The evolution of initial functions close to these NHFPS can be understood by center manifold theory \cite{hale1993,diekmann1995}.
The center manifold is invariant under the dynamics and is tangential to the center subspace spanned by the eigenfunctions belonging to stability exponents with zero real part.
The dynamics on the center manifold in the vicinity of the fixed point solutions $x_{\text{nh}}^*(t)$ is approximated up to third order by the ordinary differential equation
$\dot{y}(t)=\Theta/(\Theta+1)\,4\pi^3\mu\,y^3(t)$ with $y(t)=x(t)-x_{\text{nh}}^*(t)$ \cite{albers2024}, which, except for the prefactor $\Theta/(\Theta+1)$,
is identical to the continuous-time approximation of the dynamics of the double-sine map in the vicinity of the marginally unstable fixed points \cite{geisel1984,bel2006}.
This leads to identical residence time statistics with a diverging mean leading to anomalous subdiffusion.
For $A\rightarrow0$, the doubly laminar phases loose their plateau character and become classical laminar phases \cite{manneville1979,pomeau1980,manneville1980},
whereas chaotic laminar phases become high-dimensional through higher order generalized laminar chaos \cite{mueller2019} and merge with the high-dimensional turbulent bursts.
However, the probability that a generic solution of the DDE comes close to the NHFPS is relatively small
because of the large effective dimension of the turbulent phases which we found to be proportional to $\Theta$ as in the case of chaotic systems with constant delay \cite{farmer1982}.
Correspondingly, we numerically found that the distribution of turbulent phases is exponential with a mean value that increases exponentially with their effective dimension.
With an appropriate stochastic model \cite{albers2024}, one can show that there is a crossover time in the corresponding EASD from normal to anomalous diffusion, which increases exponentially with $\Theta$.
Therefore, for large values of $\Theta$ (large delay limit), subdiffusion might not be observable as it is the case in Fig.~\ref{fig:1}.
For small values of $\Theta$, however, we were able to observe subdiffusion even in the case of a constant delay.
For increasing amplitudes $A$ of the delay modulation, the effective dimension decreases as it is shown in Fig.~\ref{fig:5}
by means of the Kaplan-Yorke dimension calculated only from turbulent and chaotic laminar phases.
This reduction of the dimension allows the observation of anomalous subdiffusion in the large delay limit also for shorter time series.

In summary, a coexistence of doubly laminar, chaotic laminar, and turbulent phases alternating within single time series was found in a typical class of time-delayed systems,
where unstable nonhyperbolic fixed point solutions play a central role.
The non-hyperbolicity is triggered by a cubic ($z=3$) inflection point in the (retarded) nonlinearity, which is typical for systems at bifurcation points.
This means that our results about chaotic subdiffusion, weak chaos, and weak ergodicity breaking generalize to many other systems with $z=3$, but also to the much wider class of systems with $z>2$.
Finally, we emphasize that in such systems crossover times may depend exponentially on system parameters, which can cause anomalous behavior to become unobservable on a short time scale.

\begin{acknowledgments}
The authors gratefully acknowledge funding by the Deutsche Forschungsgemeinschaft (DFG, German Research Foundation) - 438881351;  456546951.
\end{acknowledgments}

\bibliography{references.bib}

\end{document}